\documentclass[cits]{PoS}
\rightline{\parbox{4cm}{\large\rm
Edinburgh 2011/35
}}

\usepackage{amsmath,amsthm,amsfonts,amssymb}
\usepackage[utf8]{inputenc}
\usepackage{subfig}

\renewcommand{\href}[1]{}
\newcommand{\path}[1]{#1}

\title{Accelerating QDP++/Chroma on GPUs}

\ShortTitle{Accelerating QDP++/Chroma on GPUs}

\author{\speaker{Frank~Winter}\\
         School of Physics and Astronomy,
                    University of Edinburgh,
                    Edinburgh EH9 3JZ, UK \\
        E-mail: \email{frank.winter@ed.ac.uk}}

\abstract{
Extensions to the C++ implementation of the QCD Data Parallel
Interface are provided enabling acceleration of expression evaluation
on NVIDIA GPUs. Single expressions are off-loaded to the device memory
and execution domain leveraging the Portable Expression Template
Engine and using Just-in-Time compilation techniques. Memory
management is automated by a software implementation of a cache
controlling the GPU's memory. Interoperability with existing Krylov
space solvers is demonstrated and special attention is paid on 'Chroma
readiness'. Non-kernel routines in lattice QCD calculations typically
not subject of hand-tuned optimisations are accelerated which can
reduce the effects otherwise suffered from Amdahl's Law.
}

\FullConference{
The XXIX International Symposium on Lattice Field Theory, Lattice 2011\\
July 10-16, 2011\\
Squaw Valley, Lake Tahoe, California}

\begin{document}

\section{Introduction}

Graphic Processing Units (GPUs) are getting increasingly important
as target architectures in scientific High Performance Computing
(HPC). The massively parallel architecture for floating-point
arithmetic together with a very high bandwidth to device-local memory
make GPUs interesting not only for compute-intensive but also for
data-intensive applications.

NVIDIA established the Compute Unified Device Architecture (CUDA) as a
parallel computing architecture controlling and making use of the
compute power of their GPUs. Now in its 4th major software iteration
mature support of most of the C++ language features (like templates)
is provided making it an interesting platform also for software
projects employing meta-programming techniques.

Within the U.S. SciDAC initiative a unified programming environment was
developed -- the QCD Application Programming Interface (API)
\cite{scidac}. 
This API enables lattice QCD scientists to implement
portable software achieving a high level of software sustainability.
Part of this API is QDP++, the C++ implementation of the QCD Data
Parallel Interface, which provides data parallel types and
expressions suitable for lattice field theory.
The very successful lattice QCD software suite Chroma builds on top
of QDP++ where implementations for a large range of hardware
architectures exist \cite{Edwards:2004sx}.
High efficiency is provided through a flexible
interface that permits specialised compute kernels to be applied
\cite{Boyle2005}.
QDP++ makes substantial use of
template meta-programming techniques to provide Domain Specific
Language (DSL) 
abstractions for this problem domain. Through usage of the Portable
Expression Templates Engine (PETE) QDP++ provides user expressions
that look similar to their mathematical counterparts.

PETE is a portable implementation of the Expression Template (ET)
technique \cite{pete, exprtemplates} -- a technique that can be used
to implement vector expressions without relying on vector sized
temporaries. PETE's portability concepts include abstractions for a
flexible return type system and user defined expression tree
traversals. However, PETE and so QDP++ do not support heterogeneous
multicore architectures with separate memory and execution domains.

This work extends QDP++/Chroma to make use of NVIDIA's CUDA as the
target architecture for expression evaluation.
A single expression is off-loaded to the device memory and execution
domain by dynamically generating a CUDA kernel and using Just-in-Time
(JIT) compilation techniques. Special attention is paid on 'Chroma
readiness' meaning that a successful build of Chroma on top of the
extended QDP++ is possible.

\begin{figure}[t]
\begin{center}
\includegraphics[width=0.5\columnwidth]{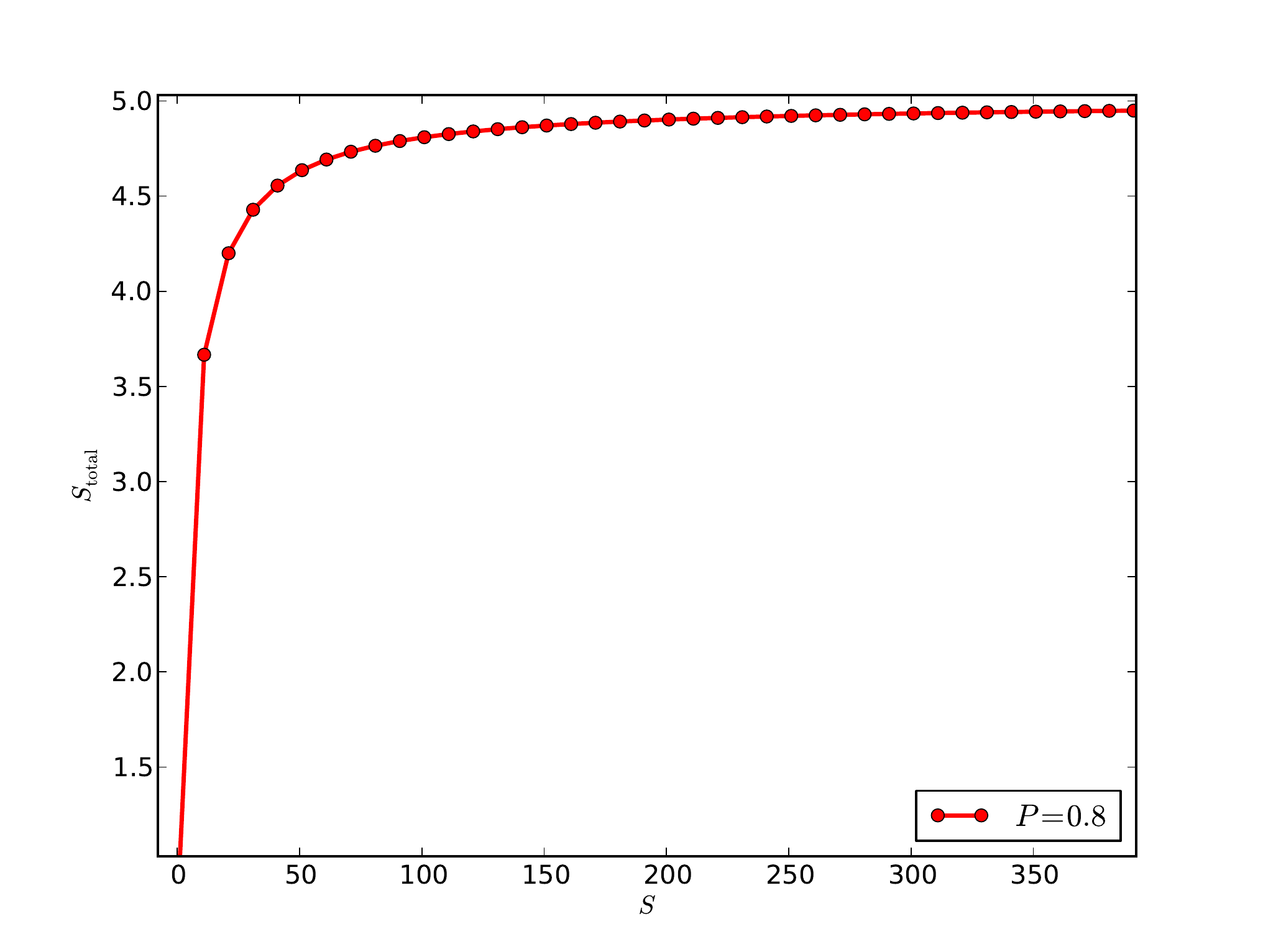}\hfill
\end{center}
\vspace*{-3mm}
\caption{\label{fig:amdahl}Amdahl's Law: Total speedup factor against speed up factor for program fraction $P=0.8$.}
\end{figure}

\begin{figure}[b]
\begin{center}
\includegraphics[width=0.5\columnwidth]{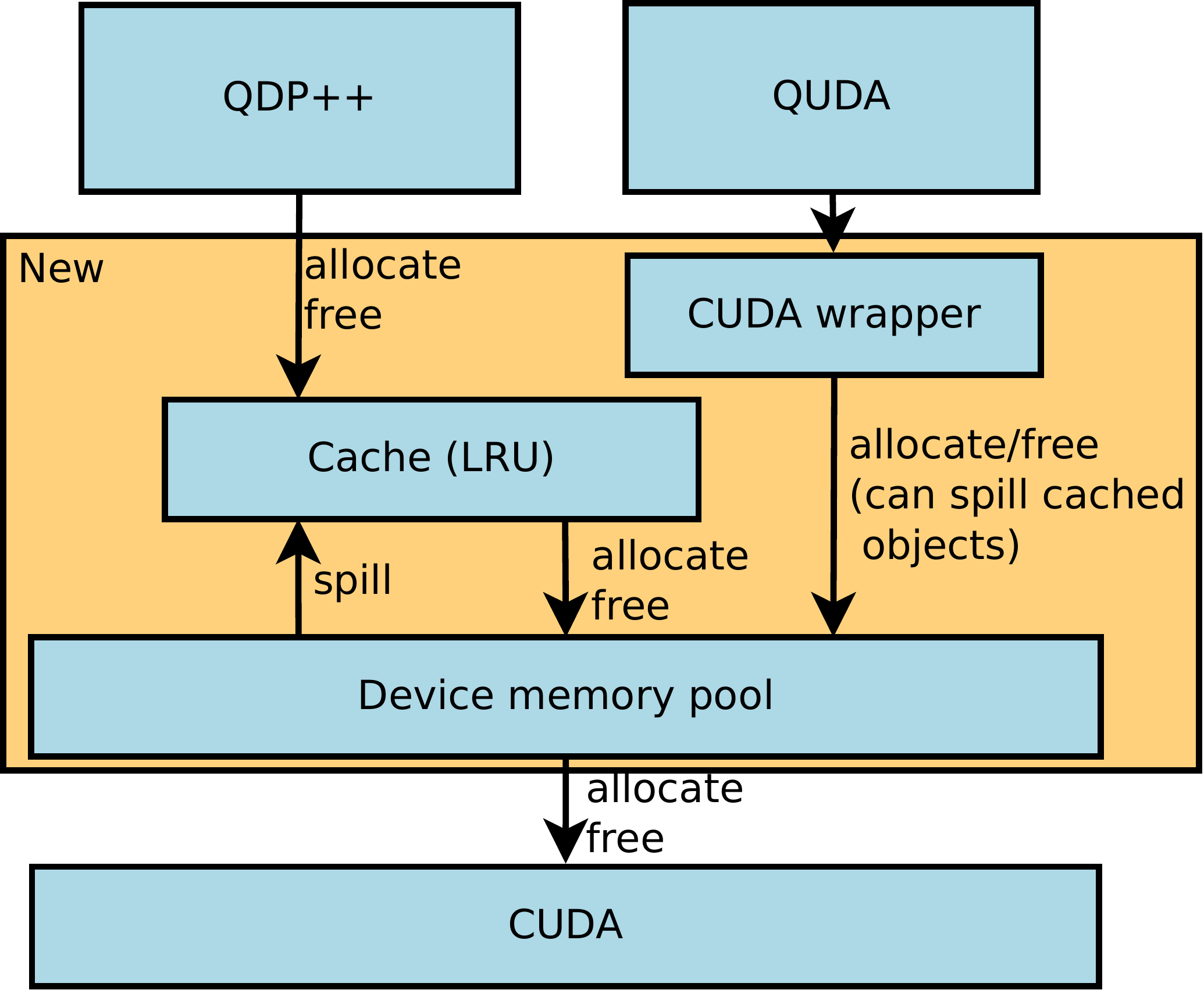}\hfill
\end{center}
\vspace*{-3mm}
\caption{\label{fig:qmem}
Software cache controlling the device memory pool and QUDA integration.}
\end{figure}

\begin{figure}[t]
\begin{center}
\includegraphics[width=0.75\columnwidth]{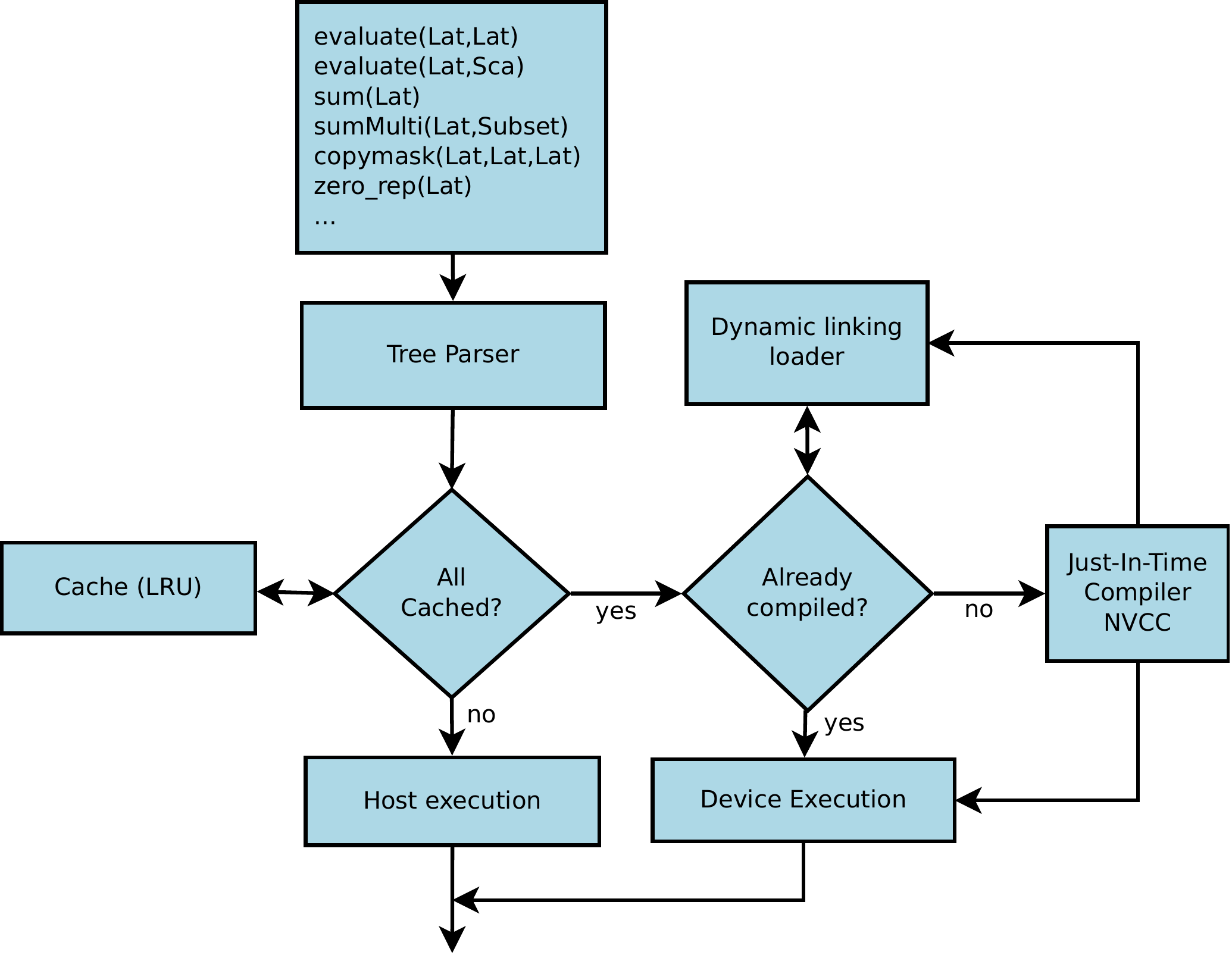}\hfill
\end{center}
\vspace*{-3mm}
\caption{\label{fig:jit}JIT control flow. The tree parser generates at
runtime GPU code and a list of lattice objects. The cache is queried
for availability of all objects on the device. JIT compilation and
device execution is triggered accordingly or alternatively host
execution.}
\end{figure}

\begin{figure}[t]
  \begin{center}
    \subfloat[A][GeForce~GTX~480, lattice size $16^3\times32$,
      $\kappa=0.13420$, $\beta=5.20$, (SP):
      Left bar: MI,SRC,SNK,HAD(CPU)
      Middle bar: MI(GPU), SRC,SNK,HAD(CPU)
      Right bar: MI,SRC,SNK,HAD(GPU).
      \label{fig:bench16}]{\includegraphics[width=.5 \textwidth]{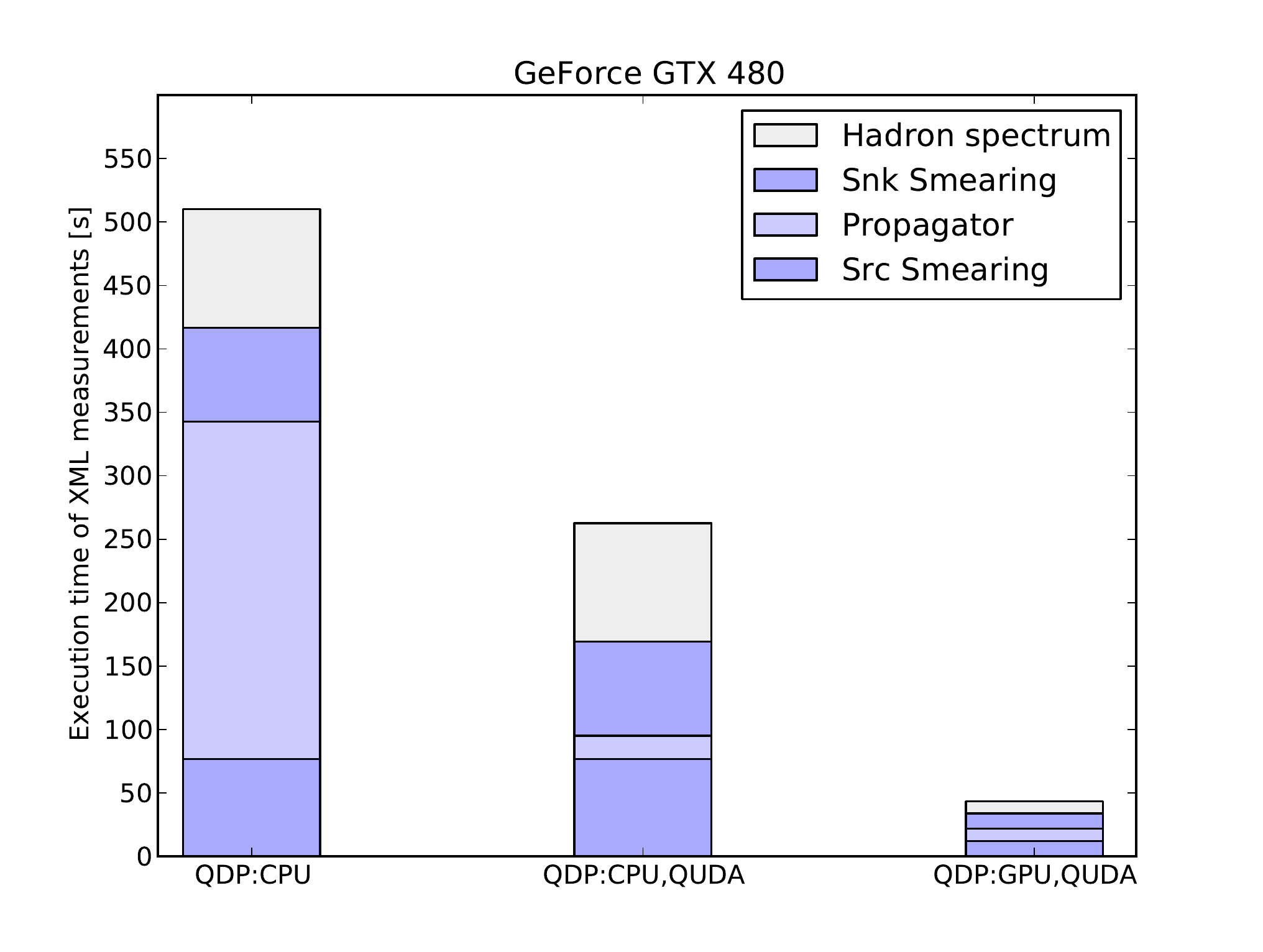}}\hspace*{3mm}
    \subfloat[A][Tesla~C2070, lattice size $24^3\times48$,
      $\kappa=0.13632$, $\beta=5.29$, (SP):
      Left bar: MI(GPU), SRC,SNK,HAD(CPU)
      Right bar: MI,SRC,SNK,HAD(GPU).
      \label{fig:bench24}]{\includegraphics[width=.5 \textwidth]{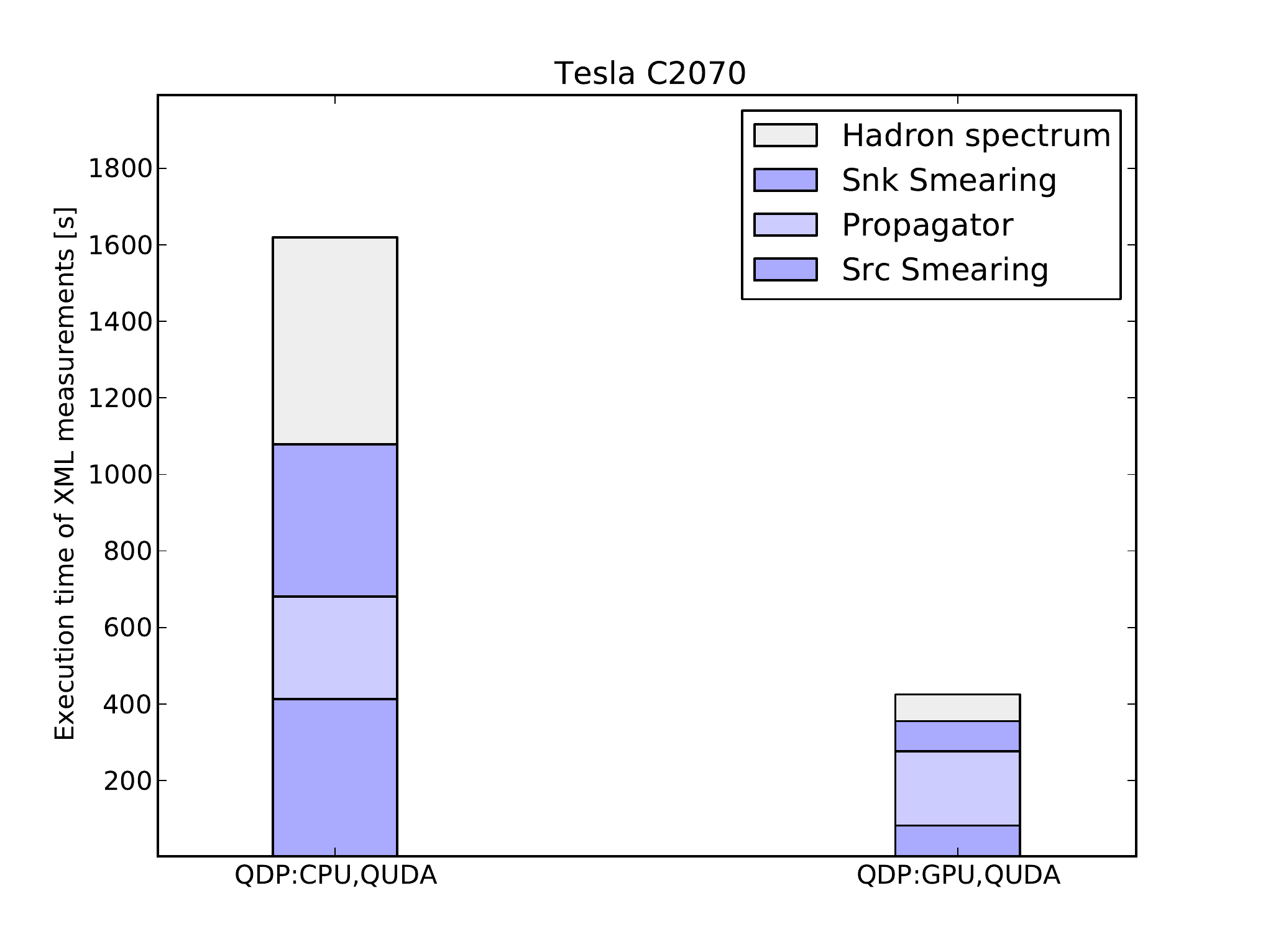}}
  \end{center}
  \vspace*{-3mm}
  \caption{
    \label{fig:bench}
    Comparison of wall-clock execution times of Chroma reference
    runs. Source smearing (SRC), Matrix inversion (MI), Sink smearing
    (SNK), Hadron spectrum (HAD).}
\end{figure}

\section{Related Work}

Efforts similar to this work are undergone at Jefferson Lab
\cite{lat11joo}. This underlines the necessity of this approach.

A similar ET reconstruction on GPUs using CUDA was previously reported
\cite{wiemann2011cuda}. Pointers to the vector's data are passed to
the CUDA kernel as function arguments and the NIVIDA compiler called
just-in-time. Memory management was not addressed and
circumvented by using the Thrust library featured by CUDA.

In previous work QDP++/Chroma was extended in a similar way targeted
to a different heterogeneous multicore architecture, the Cell 
processor and QPACE \cite{winter-phd,Nakamura:2011cd}.

\section{Lattice QCD, GPUs and Amdahl's Law}

Lattice QCD calculations divide into two parts, the generation of
background field configurations and the computation of observables on
these configurations, the so-called analysis part. The computationally
most intensive part of the analysis part is the inversion of the
fermion matrix. Although heavily dependent on the simulation
parameters the vast majority of the total amount of floating-point
operations carried out during the analysis is spent for inverting the
large sparse fermion matrix. The rest of the floating-point
operations are spent on non-kernel routines like smearing, quark
contractions, etc.

It is natural to spend most optimisation work on the inverter
part. A highly optimised library (QUDA) for the fermion matrix
inversion for NVIDIA GPUs is available \cite{Clark:2009wm,
  Babich:2010mu, quda}. These inverters provide speedup factors of over $S
\ge 30$ compared to an inversion carried out on the CPU, see a later
section for benchmark results.

However, Amdahl's law states that a program fraction $P$ subject to
acceleration with the according program part sped up by a factor $S$
gains a total speedup factor of
\begin{equation}
  S_\text{total}=\frac{1}{(1-P)+\frac{P}{S}}.
\end{equation}
For a very high speedup factor $S\rightarrow\infty$ the total speedup
factor is limited by the fraction $S_\text{total}=1/1-P$.
Fig. \ref{fig:amdahl} shows $S_\text{total}$ over $S$ for $P=0.8$.
To further increase $S_\text{total}$ one needs to increase $P$.

\section{QDP++ Extensions for GPUs}

To further increase $P$ in case of Chroma one can either implement
more hand-tuned versions of non-kernel routines or target on the
underlying library QDP++. 
Targeting on QDP++ by adding design extensions for GPU support is
advantageous since this approach results in a more general
solution. General in the sense that the user is not restricted to
specific non-kernel routines.

\subsection{Memory Management}

The bandwidth between host and device memory domain represents a
major bottleneck. Since lattice objects are typically more often
referenced than just once in a particular set of expressions
minimising these transfers can be accomplished by an implementation of
a software cache controlling the memory domain affiliation of
individual lattice objects. Provided with enough device memory
re-referencing lattice objects does not trigger transferring them
again.

Fig. \ref{fig:qmem} shows the functional principle. A pool manager
allocates at program startup time a large portion of the GPU memory
and delegates control to the cache. Upon dynamic memory allocation the
caching algorithm spills if necessary the least recently used (LRU)
object(s). This automates the memory management and application codes,
e.g. Chroma, build without changes to the code.

\subsection{Just-in-Time Compilation}

The expressions are not known at library development time. A dynamic
code generator is implemented using PETE's user defined expression
tree traversals (tree parser). Specialised leaf functors generate GPU
code for references to lattice objects and collect memory addresses of
involved lattice objects populating the parser list. Specialised
actions for tree nodes then rebuild the operations and the structure
of the expression.

Fig. \ref{fig:jit} shows the JIT compilation control
flow. Upon expression evaluation the tree parser generates GPU code
for the expression and the parser list containing lattice objects. The
cache is queried for availability of the lattice objects on the
device. In case all objects are cached, i.e. available on the device, the
availability of the CUDA kernel is queried via the dynamic linking
loader. If no CUDA kernel for the expression can be found JIT
compilation is triggered using NVIDIA's FrontEnd++ and the resulting
kernel is dynamically loaded. Then device execution is started.

\subsection{QUDA Integration}

Special emphasis is put on the interoperability of the memory
management via the LRU cache and QUDA. QUDA makes use of CUDA's API to
allocate device memory. Call wrappers are in place that redirect
memory allocation calls to the device memory pool manager controlled
by the LRU cache. Fig. \ref{fig:qmem} shows the interoperability of
QUDA with the device memory pool. QUDA memory allocation might first
trigger cache spills, then memory allocation takes place via the pool
manager.
This permits QUDA and QDP++ sharing the same device memory pool and
thus avoids the necessity to temporarily suspend QDP++ operation on
the device during propagator calculation. As a side effect this speeds
up the propagator calculation since residual calculation and
solution reconstruction are implemented using QDP++.

\section{Benchmark Results}

Chroma was used for the benchmark measurements. Three configurations
were used: 
\begin{itemize}
\item QDP++ CPU, no QUDA
\item QDP++ CPU, with QUDA
\item QDP++ GPU, with QUDA
\end{itemize}
For each of these configurations the same set of calculations was
carried out: Source creation, smearing, Propagator calculation
(Wilson-Clover), sink smearing and hadron spectrum (mesonic and
baryonic). Each of the individual calculation was timed separately.

Fig. \ref{fig:bench} shows the comparison for the different 
configurations. Shown are the individual execution times (wall-clock)
for the different Chroma measurements.

Fig. \ref{fig:bench16} refers to benchmark runs carried out on a
NVIDIA GeForce GTX 480 (1.5GB device memory, consumer product).
The left most bar represents the
configurations with all calculations carried out on the CPU (Intel
Xeon CPU, 4 cores, 2.27GHz). This
result is to be compared to the middle bar showing the execution time
of the configuration with QUDA, i.e. the matrix inversion uses the GPU
and all remaining calculations carried out on the CPU. Even the
speedup factor for the matrix inversion is about $S\approx 30$ an
overall speedup factor of only roughly $S_\text{total} \approx 2$ is
measured -- non-kernel routines (smearing and hadron spectrum) start
to dominate the total execution time. Note, however, that a rather
large quark mass was chosen. The right bar shows the execution time
when using the QDP++ with GPU evaluation. The remaining parts are
accelerated and the execution time is significantly reduced. Also the
propagator part achieved an additional speedup since residual
calculation and solution reconstruction is implemented using
QDP++. This leads to an overall speedup factor of more than
$S_\text{total} \ge 10$.

Fig. \ref{fig:bench24} shows the benchmark results carried out on a
NVIDIA Tesla C2070 (6GB device memory). In
this run the quark mass was chosen to be smaller and the lattice size
to be larger. The Chroma configuration with all parts carried out on
the CPU was not measured. The left bar shows the execution time of
using QUDA plus the remainder executed on the CPU. The right bar shows
the according time for all parts executed on the GPU. Again non-kernel
routines show a significant speedup factor.

\section{Conclusion and Outlook}

Acceleration of QDP++ expression evaluation was achieved using a
single GPU. The results look encouraging to further advance this
work. For now device memory accesses are not coalesced resulting
in a bandwidth usage of only $\approx 15-25$ GB/s ($9-15\%$ peak) on
the benchmarked devices. Initial testing with coalesced memory
accesses achieved a much higher sustained bandwidth equivalent to a
factor $\approx 6-10$ higher.
A single GPU is supported for now. Parallelisation to multiple GPUs
per host and multiple hosts forms part of future work.

\section{Code Availability}

This QDP++ implementation with extensions for GPU evaluation is
available at:

\texttt{https://github.com/fwinter/qdp}

A modified QUDA including call wrappers to QDP++ memory management is
available at:

\texttt{https://github.com/fwinter/quda}

\section{Acknowledgement}

This research was supported by the Research Executive Agency (REA) of the European Union under Grant Agreement number PITN-GA-2009-238353 (ITN STRONGnet).

\bibliography{lattice}

\end{document}